\documentclass[epj]{svjour}
\usepackage{graphics}
\usepackage{hyperref}
\usepackage{amsmath,amssymb}
\usepackage{multirow}
\usepackage{tikz} 
\usepackage{stackrel}
\usepackage{orcidlink}
\usepackage{widetext}
\usetikzlibrary{matrix} 
\hypersetup{
    colorlinks=true,
    linkcolor=blue,
    filecolor=blue,      
    urlcolor=blue,
    citecolor=blue}
\begin{document}
\title{Spherically Symmetric Geometrodynamics in Jordan and Einstein frames}
\author{Matteo Galaverni\orcidlink{0000-0002-5247-9733}\inst{1,2,3} \and Gabriele Gionti, S.J.\orcidlink{0000-0002-0424-0648}\inst{1,4,5}
}                     
%
%
\institute{
Specola Vaticana (Vatican Observatory), V-00120 Vatican City, Vatican City State \and
INAF/OAS Bologna, via Gobetti 101, I-40129 Bologna, Italy \and
INFN, Sezione di Bologna, via Irnerio 46, 40126 Bologna, Italy \and
Vatican Observatory Research Group, Steward Observatory, The University Of Arizona,
933 North Cherry Avenue, Tucson, Arizona 85721, USA  \and 
INFN, Laboratori Nazionali di Frascati, Via E. Fermi 40, I-00044 Frascati, Italy 
}
\mail{ggionti@specola.va}
\date{Received: date / Revised version: date}
%
\abstract{
Spherically symmetric geometrodynamics is studied for scalar-tensor theory and Einstein General Relativity minimally coupled to a scalar field. 
We discussed the importance of boundary terms for non-compact space-like foliations and derived the equations of motion in the Hamiltonian formalism both in the Jordan and Einstein frames. 
On the reduced phase space obtained by gauge-fixing the lapse and the radial shift functions, the two frames are connected through a Hamiltonian canonical transformation. 
We discussed the effects of the singularity of the Hamiltonian canonical transformation connecting Jordan and Einstein frames for two static solutions (Fisher, Janis, Newman and Winicour solution in the Einstein frame and Bocharova-Bronnikov-Melnikov-Bekenstein black hole solution in the Jordan frame).
\PACS{
      {PACS-key}{discribing text of that key}   \and
      {PACS-key}{discribing text of that key}
     } 
} 
\maketitle
\section{Introduction}

It is fairly well known that early attempts to quantize Einstein General Relativity stem from the general procedure to cast the theory into the Hamiltonian formalism first and then to apply all the procedure to pass from the classical to the quantum world through the so called {\it Dirac's map} \cite{Dirac1930-DIRTPO} (canonical quantization). The main works on the Hamiltonian formalism in Einstein General Relativity were pursued by Bergmann \cite{Bergmann:1949zz,RevModPhys21480}, Dirac himself \cite{dirac1966}, Arnowitt, Deser and Misner \cite{ADM}. 
Bergmann and Dirac developed a formalism to cast singular Lagrangians into Hamiltonian theory. These Lagrangians, because of the local symmetries in gauge-theories and Einstein's General Relativity, have their Hessian determinant equal to zero. 
Then, it is not, strictly speaking, possible to define the momenta associated to their corresponding velocities. Bergmann and Dirac also developed a method for defining momenta in these cases. 
Arnowitt, Deser and Misner (ADM) focused on the geometric splitting of space-time into space and time (3+1 decomposition), so that it can be used as basis to construct all the Hamiltonian formalism on it.

Dicke in a seminal article \cite{Dicke} stressed that physics is invariant under redefinition of the units of measurement (see also \cite{Faraoni2006}). 
He noticed that invariance under a local rescaling of the units of measurement 
implies invariance under Weyl (conformal) transformation of the metric coefficients. 
This observation is the starting point of the Jordan and Einstein frames topic. 
The Einstein frame (EF) is always the frame where the action is the Einstein-Hilbert action minimally coupled to a scalar field. Performing a Weyl (conformal) transformation of the metric coefficients and using a conformal factor functionally dependent by the scalar field, the final result of the transformation is, in general, a scalar tensor theory. This latter result defines the Jordan frame (JF). 
Following Dicke \cite{Dicke}, physical quantities can be calculated either in JF or in the EF;
they are related by a power of the Weyl (conformal) factor according to their dimensions in natural units 
(e.g. see \cite{Chiba:2013mha,Francfort:2019ynz}). 
A particular scalar-tensor theory in the JF, e.g. Brans-Dicke theory, is mapped by a specific Weyl (conformal) transformation - functionally dependent by the scalar field - into a Einstein-Hilbert action in the EF. 
In the literature \cite{Deruelle2010h}, it is generically assumed that if we have a solution of the equations of motion in the JF, then the corresponding Weyl (conformal) transformed counterpart is solution of the equations of motion in the EF. In principle, there is no ``a priori'' reason why this should be true. The underlying reasoning is that the passage from JF to EF is just a field redefinition. 
Then the solutions of the equations of motion in the JF can be mapped, by the transformation from JF to EF, into solutions of the equations of motion in the EF and vice versa. 
A way to verify this is to pass to the Hamiltonian formalism and check whether the transformation from JF to the EF is an Hamiltonian canonical transformation. In the past, either it was assumed that this transformation was Hamiltonian canonical \cite{Garay1992} or this statement was only partially proved \cite{Deruelle2009}.

In a series of articles \cite{Gionti2021,Galaverni:2021xhd,Galaverni:2021jcy,GiontiSJ:2023tgx}, we studied the question of equivalence between JF and EF. 
We started with the Brans-Dicke theory, the simplest scalar-tensor theory. 
We passed to the corresponding Hamiltonian theory; 
we performed the Dirac's constraint analysis in JF and in EF separately. 
We noticed that the Hamiltonian transformation from the JF to the EF is not Hamiltonian canonical on the extended phase-space where the Dirac's constraints are not zero \cite{Gionti2021}. 
We have studied both the cases $\omega=-\frac{3}{2}$ and $\omega\neq-\frac{3}{2}$ in the Brans-Dicke theory, in the JF, and the corresponding  theory in the EF \cite{Galaverni:2021xhd}. 
The study of the equivalence between the two frames based on a flat FLRW mini-superspace model \cite{Galaverni:2021jcy} suggested that, gauge-fixing the lapse and the shifts functions, we could obtain  that the Hamiltonian transformation between JF and EF might be Hamiltonian canonical transformation on a manifold with a reduced number of variables. 
In \cite{GiontiSJ:2023tgx} we showed that if we gauge-fix the lapse and the shifts functions and implement them as secondary constraints, then the primary first class constraints become second class and we can define Dirac's brackets, which replace the Poisson brackets. 
We derived the equations of motion using Dirac's brackets. 
Then, we impose strongly the second class constraints. 
This generates a reduced phase space. On this reduced phase space,  the Hamiltonian transformation between Jf and EF is completely Hamiltonian canonical. 
Of course, this particular type of  mathematical equivalence does not necessarily imply physical equivalence as we have discussed extensively in \cite{GiontiSJ:2023tgx} with the support of examples. 

In this paper, we focus on ADM analysis in the particular case of spherical symmetry following \cite{Kuchar:1994zk},
see also \cite{Berger:1972pg,Lund:1973zzb,CORDERO1976607,Kuchar:1994zk,Lau:1995fr,Esposito:2007xz,Christodoulakis:2012eg,Rosabal:2021fao,Rosabal:2023sbz}. 
A scalar tensor theory is analyzed in the Hamiltonian formalism, and we study, carefully, the boundary terms to be included in its action in order to make the variation of the action linear in the variation of the metric coefficients.  
We show that it is crucial to properly consider all boundary terms in order to derive the correct definitions of canonical momenta 
and obtain the Hamiltonian equations of motion 
Finally, we consider two spherically symmetric static solutions: we show that the Fisher  \cite{Fisher:1948yn}, Janis, Newman and
Winicour \cite{Janis:1968zz} solution (naked singularity) is transformed into the Bocharova-Bronnikov-Melnikov-Bekenstein solution \cite{Bocharova:1970,Bekenstein:1974sf} (black hole). 
These two physically in-equivalent solutions can be mapped one into the other since the Hamiltonian canonical transformation between the two frames is singular.

This is the plan of the essay. 
In Section~\ref{section:2} we provide a review of ADM formalism for Einstein's General Relativity and specify it for spherically symmetric systems.
The spherically symmetric ADM is then applied to the EF case in Section~\ref{section:EF} 
and to JF in Section~\ref{section:JF} for a scalar-tensor theory, with a particular attention to the boundary terms.
In Section~\ref{Sect:sol} we discuss two particular static solutions in the two frames and discuss how they are connected by conformal transformations.
We conclude in Section~\ref{Conclusions}.
Through all the paper natural units are used: $G=1=c$.

\section{ADM formalism for Einstein's General Relativity}
\label{section:2}

The aim of this Section is to provide a review of ADM formalism for Einstein's General Relativity first in the general case  (\ref{section:2:1}) and later specify it for a system with spherical symmetry (\ref{Sect:ssADM}).

\subsection{ADM formalism in general}
\label{section:2:1}

\begin{figure}[htb]
    \centering
    \includegraphics[width=0.8\linewidth]{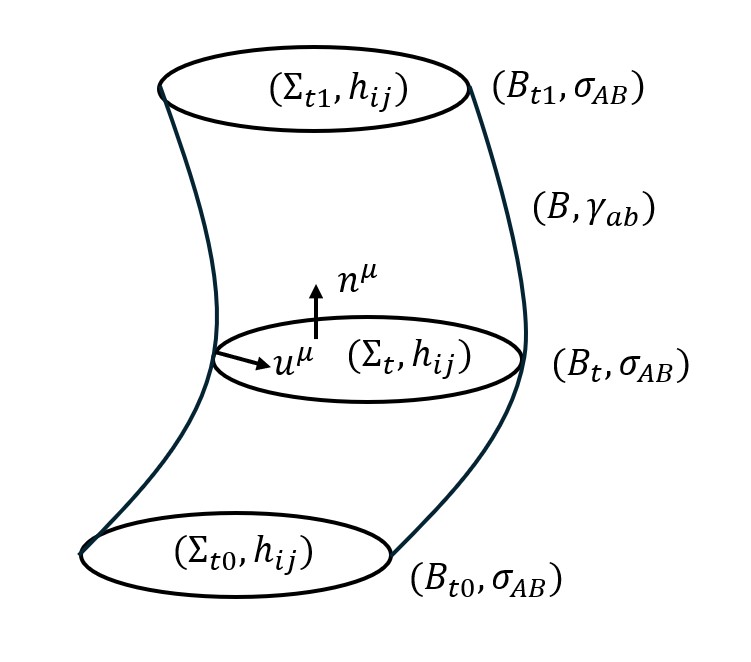}
    \caption{The 4-dimension manifold $(M,g_{\mu\nu})$ can be foliated in 3-dimensions space-like sub-manifolds at a given time $(\Sigma_t,h_{ij})$ with normal vector $n^\mu$. The time-like boundaries correspond to a 3-dimension time-like foliation  
    $(B,\gamma_{ab})$ with normal vector $u^\mu$. Note that in general the space-like and the time-like foliations are not orthogonal ($n^\mu u_\mu\neq 0$).}
    \label{fig:1}
\end{figure}

We consider a Lorentzian manifold $(M,g_{\mu\nu})$, $M$ being a differential manifold and $g_{\mu\nu}$ a Lorentian metric defined on it  with signature $(-,+,+,+)$, with a ADM $3+1$ decomposition such that the topology of $M=\mathbb{R} \times \Sigma_t$. 
Here $\mathbb{R}$ is the one-dimensional time direction and $\Sigma_t$ is a non-compact three-dimensional space-like surface as in  \cite{Kuchar:1994zk}. 
Foliation of manifold $M=\mathbb{R} \times \Sigma_t$ is visualized in Fig.~\ref{fig:1}. 

The boundary of $M$ is nothing but $\partial M=\Sigma_{t_0} \cup \Sigma_{t_1}\cup B$. 
The sub-manifolds $\Sigma_{t_0}$ and $\Sigma_{t_1}$ are two space-like leaves of the $3+1$ decomposition of $M$, $n^{\mu}$ is the time-like normal vector field to them and $h_{ij}$ is the three-dimensional metric tensor defined on these space-like leaves. $B$ is the time-like boundary of $M$, whose normal vector field $u^{\mu}$ is space-like and $\gamma_{ab}$ is the three-dimensional metric tensor defined on $B$. $B_t$ is the two dimensional manifold, intersection of the space-like leaves $\Sigma_t$ with the time-like boundary $B$; the two dimensional metric tensor on this sub-manifold is $\sigma_{AB}$. $K_{ij}$ is three-dimensional extrinsic curvature on $\Sigma_t$, $\Theta_{ab}$ is the three-dimensional extrinsic curvature on $B$ and ${}^{(2)}k_{AB}$ the two-dimensional extrinsic curvature on $B_t$. 
See Tab.~\ref{tab:symbols} for a summary of the symbols used.

\begin{table*}[ht]
    \begin{tabular*}{\textwidth}{@{\extracolsep{\fill}}|l|c|c|c|c|@{}}
        \hline
        (sub)manifold  & $M$ & $\Sigma_t$ & $B$ & $B_t$ \\
        \hline
        \hline
         Dimension & 4 & 3 & 3 & 2 \\
         \hline
          Indices & $\{\mu,\nu\}=\{t,r,\theta,\varphi\}$ & $\{i,j\}=\{r,\theta,\varphi\}$ & $\{a,b\}=\{t,\theta,\varphi\}$ & $\{A,B\}=\{\theta,\varphi\}$ \\  
         \hline
         Normal vector & & $n_\mu$ & $u_\mu$ & $r_\mu$ \\
         \hline
         Metric & $g_{\mu\nu}$ & $h_{\mu\nu}$  & $\gamma_{\mu\nu}$ & $\sigma_{\mu\nu}$  \\
         \hline
         Covariant derivative & $\nabla_\mu$  & $D_\mu$ &  & \\
         \hline
         Ricci scalar & ${}^{(4)}R$ & ${}^{(3)}R$ & & \\
         \hline
         Extrinsic curvature &  & $K_{\mu\nu}=-h_\mu^\rho\nabla_\rho n_\nu$  & $\Theta_{\mu\nu}=-\gamma_\mu^\rho\nabla_\rho u_\nu$ & ${}^{(2)}k_{\mu\nu}=-\sigma_\mu^\rho  D_\rho u_\nu$ \\
         \hline
         Trace of the extr. curvature &  & $K$  & $\Theta$ &${}^{(2)}k$ \\
         \hline
    \end{tabular*}
    \caption{Summary of the symbols used in this paper; see also Tab. 1 of  \cite{Hawking:1996ww} and Tab.~4.1 of \cite{Poisson:2009pwt}} for similar notations.
    \label{tab:symbols}
    \end{table*}

The Einstein-Hilbert action functional with boundary terms which make it linear in the variation of the metric tensor is  \cite{York:1986lje,Lau:1995fr,Hayward:1992ix,Hayward:1993my,Hawking:1996ww,Dyer,Rosabal:2021fao,Reyes:2023sgk}:
\begin{eqnarray}
S &=&\frac{1}{16\pi}
\int_{{M}}d^{4}x \sqrt{-g}\; {}^{(4)}R +\frac{1}{8\pi }\int_{\Sigma_{t_1}}d^{3}x\sqrt{h}K \nonumber\\ 
&&-\frac{1}{8\pi }\int_{\Sigma_{t_0}}d^{3}x\sqrt{h}K 
- \frac{1}{8\pi }\int_{B}d^{3}x\sqrt{-\gamma}\,\Theta \nonumber\\
&& - \frac{1}{8\pi }\int_{B_{t_0}} d^2x \sqrt{\sigma} \, \mathrm{arcsinh} (n^\mu u_\mu)  \nonumber\\
 &&+\frac{1}{8\pi }\int_{B_{t_1}} d^2x \sqrt{\sigma} \,\mathrm{arcsinh} (n^\mu u_\mu).
 \label{tutta}
\end{eqnarray}

We can use the following $3+1$ ADM decomposition \cite{Hawking:1996ww} of the trace of the Ricci tensor ${}^{(4)}R$:
\begin{equation}
{}^{(4)}R={}^{(3)}R+K_{ij}K^{ij}-K^{2}+2\nabla_{\mu}\left(-n^{\mu}K-a^{\mu} \right)\, ,
\label{covo}
\end{equation}
where $n^{\mu}$ is the unit time-like normal to the space-like surface $\Sigma$, and $a^{\mu}\equiv n^{\nu} \nabla_{\nu}n^{\mu}$;
see Appendix~\ref{App:1} for an equivalent way to decompose the trace of the Ricci tensor.

Applying Stokes' theorem, in the case of space-like and time like boundary surfaces \cite{Poisson:2009pwt}, we get:
\begin{eqnarray}
S&=&\frac{1}{16\pi }
\int dt \int_{\Sigma_t}d^{3}x
\left(\sqrt{h}N({}^{(3)}R+K_{ij}K^{ij}-K^{2})\right)\nonumber\\
&&-\frac{1}{8\pi  }\int_{B}d^{3}x\sqrt{-\gamma}\,\left(\Theta+u_{\mu}n^{\mu}K+u_{\mu}a^{\mu} \right) \nonumber\\
&&-\frac{1}{8\pi }\int_{B_{t_0}} d^2x \sqrt{\sigma} \, \mathrm{arcsinh} (n^\mu u_\mu) \nonumber\\
&& +\frac{1}{8\pi }\int_{B_{t_1}} d^2x \sqrt{\sigma} \,\mathrm{arcsinh} (n^\mu u_\mu).
 \label{aritutta}
\end{eqnarray}
In the second line of this last formula the quantity in round parenthesis of the integrand is proportional to the extrinsic curvature ${}^{(2)}k$ on two-dimensional boundary $B_t$ \cite{Hawking:1996ww}.

This action is well defined for spatial compact geometries, but it is divergent for non-compact ones \cite{Hawking:1995fd}. 
In order to define the action for non-compact geometries, we choose a background geometry $g_0$ and assume that $g_0$ is a asymptotically flat static solution of the equations of motion. We define the {\em physical action} $S^{\mathrm{PHYS}}$ as: 
\begin{equation}
S^{\mathrm{PHYS}}(g)\equiv S(g)-S(g_0)\,, 
\label{rinorma2}
\end{equation}
so that the physical action of the reference background is zero. $S^{\mathrm{PHYS}}$ is defined for a class of metrics $g$ that asymptotically approach $g_0$. The metric $g$ should approach asymptotically to $g_0$, in the sense that if we consider a boundary near the infinity $\Sigma_{\infty}$, the metric $g$ induces on this boundary the same fields as $g_0$.

Following the standard procedure described in \cite{Hawking:1995fd,Hawking:1996ww,Poisson:2009pwt,Rosabal:2021fao} the physical action can be written as:
\begin{eqnarray}
S^{\mathrm{PHYS}} &=&\frac{1}{16\pi }
\int_{t_0}^{t_1} dt \int_{\Sigma_t} d^{3}x N \sqrt{h} \left({}^{(3)}R+K_{ij} K^{ij}-K^2\right) \nonumber\\
&& -\frac{1}{8\pi }\int_{t_0}^{t_1}dt \int_{B_t}d^{2}x N \sqrt{\sigma} \left({}^{(2)}k-{}^{(2)}k_0 \right)\,. 
\label{phys:action:1}
\end{eqnarray}
The terms due to the non orthogonality of the boundaries (tilting terms) are proportional to $\mathrm{arcsinh} (n^\mu u_\mu)$
and they cancel when the background action is subtracted,
as extensively discussed in \cite{Hawking:1996ww}.

The momentum conjugate to $h_{ij}$ does not get contribution from the boundary terms, therefore we get:
\begin{eqnarray}
\pi_{ij}\equiv\frac{\partial\mathcal{L}_{\mathrm{PHYS}}}{\partial \dot{h}_{ij}}=\frac{\partial K_{lm}}{\partial\dot{h}_{ij}}\frac{\partial\mathcal{L}_{\mathrm{PHYS}}}{\partial K_{lm}}=-\frac{\sqrt{h}}{16 \pi } \left(K^{ij}-K h^{ij}\right)\,.
\end{eqnarray}
The total Hamiltonian is obtained integrating $\mathcal{H}_{ADM} \equiv \pi^{ij} \dot{h}_{ij}-\mathcal{L}_{P}$ over $\Sigma_t$ and adding the boundary terms following the standard procedure \cite{Hawking:1996ww,Poisson:2009pwt,Gionti2021,Blau}:
\begin{eqnarray}
H_{ADM}&=&\int_{\Sigma_t} d^3x \left( N \mathcal{H}+N_i \mathcal{H}^i \right)\nonumber\\
&&+2 \int_{B_t}d^2x \sqrt{\sigma} \left(N_i \frac{\pi^{ij}}{\sqrt{h}}r_j-\left. N_i \frac{\pi^{ij}}{\sqrt{h}}r_j\right|_{0}\right)\nonumber\\
&&-\frac{1}{8\pi }\int_{B_t}d^2x\sqrt{\sigma}  \left({}^{(2)}k-{}^{(2)}k_0 \right)\,,
\label{Ham:adm1}    
\end{eqnarray}
where the density of the Hamiltonian constraint is:
\begin{equation}
\mathcal{H}\equiv\sqrt{h}\left[\frac{1}{16\pi }{}^{(3)}R-\frac{16\pi }{h}\left(\pi_{ij}\pi^{ij}-\frac{\pi_h^2}{2}\right) \right]\,,  
\label{H:const:0}
\end{equation}
and the momentum constraint is:
\begin{equation}
\mathcal{H}^i\equiv  -2 \sqrt{h} D_j \left(\frac{\pi^{ij}}{\sqrt{h}} \right)\,.
\label{Hi:const:0}
\end{equation}
Since the background metric $g_0$ is solution we have contribution only from the metric $g$ in the first line of \eqref{Ham:adm1}, on the contrary it is not always possible to ensure that momentum density vanishes for the background solution \cite{Hawking:1996ww}.

\subsection{ADM formalism for spherically symmetric systems}
\label{Sect:ssADM}

From here one we assume that $\Sigma_t$ has a spherical symmetry, with topology $\Sigma_t = \mathbb{R}\times S^{2}$.
The line element on a fixed sub-manifold $\Sigma_t$ with metric $h_{ij}$ depends only on two functions $\Lambda(r)$ and $R(r)$ \cite{Berger:1972pg,Kuchar:1994zk}:
\begin{equation}
h_{ij}dx^i dx^j={\Lambda}^{2}dr^{2}+R^{2}d{\Omega}^{2}\,\,
\label{spherical0}
\end{equation}
where $d\Omega$ is  the line element on the unit sphere, $d\Omega^2\equiv d\theta^2+\sin^2\theta d\varphi^2$.
A point $x \in \Sigma_t$ has coordinates $x^{i}=(r,\theta,\varphi)$, with $r\in \left]-\infty,+\infty \right[\,$, 
$\theta\in \left[0,\pi\right[$ and $\varphi\in \left[0,2\pi\right[\,\textbf{}$.
We take $\Lambda(r)>0$, being $\Lambda(r) dr$ the radial line element oriented from left infinity ($r\to-\infty$) to right infinity ($r\to+\infty$).
We also assume $R(r)>0$: $R(r)$ is the radius of a two sphere of radius $r=$const., see \cite{Kuchar:1994zk}. 

The space-time is  foliated in spherically symmetric leaves $\Sigma_t$, 
labeled leaves with a time parameter $t\in\mathbb{R}$.
The line element on the four-dimensional Lorentzian manifold $(M,g)$ is \cite{Lund:1973zzb,Kuchar:1994zk,Lau:1995fr,Friedman:1997fu,Esposito:2007xz,Loll:2017utf}: 
\begin{eqnarray}
    g_{\mu\nu}dx^\mu dx^\nu&=& - N^2 dt^2 
      +  \Lambda^2 \left(dr+N^r dt \right)^2  + R^2 d\Omega^2 \nonumber \\
    &=& - \left(N^2-\Lambda^2 (N^r)^2 \right) dt^2
      + 2 \Lambda^2 N^r dt dr\nonumber\\ 
      &&+ \Lambda^2 dr^2
      + R^2 d\Omega^2 \,, 
\label{def:ADMss}
\end{eqnarray}
therefore the metric coefficients $g_{\mu\nu}$ are: 
\begin{eqnarray}
g_{\mu\nu} = \left( 
\begin{array}{cccc}
  - \left(N^2-\Lambda^2 (N^r)^2 \right) & \Lambda^2 N^r & 0 & 0\\
  \Lambda^2 N^r & \Lambda^2  &  0 & 0\\
  0 & 0 & R^2 & 0 \\
  0 & 0  & 0 & R^2 \sin^2\theta
\end{array}
\right)\,.
\end{eqnarray}
$N$ is the lapse function and $N^{r}$ is the only shift function, related to the variable $r$, due to the spherical symmetry. They are functions of $t$ and $r$: $N=N (t,r), N^{r}=N^r(t,r)$ as $\Lambda=\Lambda (t,r)$ and $R=R(t,r)$.

The line element (\ref{spherical0})  implies that the three-dimensional scalar curvature ${}^{(3)}R[h]$ is: 
\begin{equation}
 {}^{(3)}R[ h ] = -4 \frac{R''}{\Lambda ^{2} R} + 4\frac{\Lambda' R'}{\Lambda ^{3} R} 
 - 2\frac{R'^{2}}{ \Lambda ^{2} R^{2}}  + \frac{2}{R^2} \, .
\label{eq:R}
\end{equation}
where $^\prime$ denotes derivative respect to $r$.

The normal vector to three dimensional surface $\Sigma_t$ is \cite{Hawking:1996ww,Poisson:2009pwt}:
\begin{equation}
n_\mu=(-N, 0,0,0)\,,
\end{equation}
the induced metric on space-like surface $\Sigma_t$, see \cite[p. 102]{Menotti2017} and \cite[p. 1564]{Lau:1995fr} is:
\begin{equation}
h_{\mu\nu}=g_{\mu\nu}+n_\mu n_\nu\,,    
\end{equation}
Instead the normal vector to the three dimensional surface $B$ is:
\begin{equation}
u_{\mu}=(0, \frac{\Lambda N}{\sqrt{N^2-(\Lambda N^r)^2}},0,0)\,,
\end{equation}
therefore, the induced metric on the  time-like sub-manifold $B$ is:
\begin{equation}
\gamma_{\mu\nu}=g_{\mu\nu}-u_\mu u_\nu\,.    
\end{equation}
Note that for the spherically symmetric metric \eqref{def:ADMss} considered here the boundaries surfaces $\Sigma_{t_0}$ ($\Sigma_{t_1}$) and $B$ are not orthogonal. The angle between the normals is:
\begin{equation}
n^\mu u_\mu=- \frac{\Lambda N^r}{\sqrt{N^2-(\Lambda N^r)^2}}\,,    
\end{equation}
however, as we mentioned in the previous subsection, the tilting terms does not contribute to the dynamic of the system.

The extrinsic curvature $K_{ij}$ on each leave $\Sigma_t$ of the foliation $\mathbb{R}\times\Sigma_t $ is \cite{Esposito1992,Gionti2021}:
\begin{equation}
    K_{ij}=\frac{1}{2N}\left(-{\dot{h}}_{ij}+D_{i}N_j+D_{j}N_i\right)\,,
    \label{estro}
\end{equation}
and for this spherical symmetric case $K_{ij}$ is diagonal \cite{Kuchar:1994zk}: 
\begin{eqnarray}
\label{eq:K} 
&&K_{rr}  = - \frac{\Lambda}{N} \left( \dot{\Lambda} - ( \Lambda
N^{r} \big)' \right) \,,\;\\  
&&K_{\theta \theta}  =  - \frac{R}{N} \Big( \dot{R} - R' N^{r} \Big) \,,\;
K_{\varphi \varphi} =
\sin ^{2} \theta \, K_{\theta \theta}\, .     \nonumber
\end{eqnarray}
where $\dot{}$ denotes derivative respect to time, the trace $K$ of the extrinsic curvature is:
\begin{eqnarray}
K&\equiv& K _{ij} h^{ij}\nonumber\\
&=&\frac{1}{N} \left[\frac{-\dot{\Lambda}+(\Lambda N^r)^\prime}{\Lambda}+\frac{2(-\dot{R}+R^\prime N^r)}{R}\right]\,. 
\label{eq:K_trace}
\end{eqnarray}
We proceed in analogous way also for the two dimensional extrinsic curvature on $B_t$ \cite{Poisson:2009pwt,Rosabal:2021fao}:
\begin{equation}
{}^{(2)}k_{AB}= \frac{1}{2\Lambda}\Big(\partial_r \sigma_{AB} \Big)\,\,.    
\end{equation}
It has only nonzero diagonal components:
\begin{equation}
{}^{(2)}k_{\theta\theta}= \frac{R R'}{\Lambda}\,\,,\;  
{}^{(2)}k_{\varphi \varphi} = \sin ^{2} \theta \, {}^{(2)}k_{\theta \theta}\, ,   
\end{equation}
and for the trace we have:
\begin{equation}
{}^{(2)}k=\sigma^{AB}\,{}^{(2)}k_{AB}=\frac{2}{\Lambda}\left( \frac {R'}{R}\right)\,. 
\end{equation}

\begin{widetext}
\section{ADM analysis in the Einstein frame}
\label{section:EF}

In the Einstein frame our starting point is standard Einstein General Relativity plus a massless scalar field $\phi$:
\begin{eqnarray}
\mathcal{L}_{EF}=\frac{1}{16\pi} \left({}^{(4)}\widetilde{R}-\widetilde{g}^{\mu\nu}\partial_\mu\widetilde{\phi}\,\partial_\nu\widetilde{\phi} \right)\,.
\label{EF:Lagr}
\end{eqnarray}

As discussed in the previous section, see Eq. \eqref{tutta}, the corresponding action, with boundary terms, is:
\begin{eqnarray}
S_{EF} &=&\frac{1}{16\pi}
\int_{{M}}d^{4}x \sqrt{-\widetilde{g}}\; \left({}^{(4)}\widetilde{R}-\widetilde{g}^{\mu\nu}\partial_\mu\widetilde{\phi}\,\partial_\nu\widetilde{\phi}\right) +\frac{1}{8\pi}\int_{\Sigma_{t_1}}d^{3}x\sqrt{\widetilde{h}}\widetilde{K} -\frac{1}{8\pi }\int_{\Sigma_{t_0}}d^{3}x\sqrt{\widetilde{h}}\widetilde{K} \nonumber\\
&&- \frac{1}{8\pi }\int_{B}d^{3}x\sqrt{-\widetilde{\gamma}}\,\widetilde{\Theta} - \frac{1}{8\pi }\int_{B_{t_0}} d^2x \sqrt{\widetilde{\sigma}} \, \mathrm{arcsinh} (n^\mu u_\mu)  +\frac{1}{8\pi }\int_{B_{t_1}} d^2x \sqrt{\widetilde{\sigma}} \,\mathrm{arcsinh} (n^\mu u_\mu).
 \label{tutta:EF}
\end{eqnarray}
Considering the spherically symmetric ADM metric introduced in the previous subsection the bulk term of the physical action defined 
in Eq.~\eqref{phys:action:1}, integrating over $\theta$ and $\varphi$, becomes \cite{Kuchar:1994zk}:
\begin{eqnarray}
S^{\mathrm{PHYS}}_{EF}
&=& \int dt \int _{-
\infty} ^{\infty} dr \Bigg[ -\, \frac{1}{\widetilde{N}} \left( \widetilde{R}
\big( - \dot{\widetilde{\Lambda}} + (\widetilde{\Lambda} \widetilde{N}^{r})' \big) \big( - \dot{\widetilde{R}} + \widetilde{R}'
\widetilde{N}^{r} \big) + \frac{\widetilde{\Lambda}}{2}  \big( - \dot{\widetilde{R}} + \widetilde{R}' \widetilde{N}^{r} \big)
^{2} \right) \nonumber \\ & & 
+ \, \widetilde{N} \left( -\frac{\widetilde{R} \widetilde{R}''}{\widetilde{\Lambda}} 
+\frac{\widetilde{R} \widetilde{R}' \widetilde{\Lambda}'}{\widetilde{\Lambda}^{2}}
- \frac{\widetilde{R}'^{2}}{2 \widetilde{\Lambda}}  
+\frac{\widetilde{\Lambda}}{2}  \right)
+\frac{1}{4}\Big( \frac{\widetilde{\Lambda} \widetilde{R}^{2}}{\widetilde{N}}\, \big( \dot{\widetilde{\phi}}- \widetilde{N}^{r} \widetilde{\phi}' \big)^{2} 
-\frac{\widetilde{N} \widetilde{R}^2 \widetilde{\phi}'^{2}}{\widetilde{\Lambda}}  \Big)\Bigg]\,. 
\label{eq:S-lag}
\end{eqnarray}

The boundary terms of the physical action \eqref{phys:action:1} are:
\begin{eqnarray}
 -\frac{1}{8\pi}\int_{t_0}^{t_1}dt \int_{B_t}d^{2}x \widetilde{N} \sqrt{\widetilde{\sigma}} \left({}^{(2)}\widetilde{k}-{}^{(2)}\widetilde{k}_0 \right)=- \frac{1}{8\pi}\int_{t_1}^{t_2}dt \int_{B_t} d^2x \widetilde{N} \left(\frac{2\widetilde{R}\widetilde{R}^\prime}{\widetilde{\Lambda}}\sin \theta-\left.\frac{2\widetilde{R}\widetilde{R}^\prime}{\widetilde{\Lambda}}\sin \theta\right|_{0}\right)\,. 
 \nonumber
\label{eq:S-lag:bound}
\end{eqnarray}
they do not contribute to the dynamics.

On the boundary terms the integral on $B_t$ is evaluated for $r\rightarrow\pm\infty$;
if we assume the standard fall-off conditions \cite{Kuchar:1994zk,Loll:2017utf}:
\begin{eqnarray}
\lim_{r\to\pm\infty} \widetilde{\Lambda}(t,r)  &=&1+\frac{\widetilde{M}_{\pm}(t)}{|r|}+O\left(\frac{1}{|r|^{1+\varepsilon}}\right)\,,\\
\lim_{r\to\pm\infty}\widetilde{R}(t,r)&=&|r|+O\left(\frac{1}{|r|^\varepsilon}\right)\,,\\
\lim_{r\to\pm\infty}\widetilde{N}(t,r)&=&\widetilde{N}_{\pm}(t)+O\left(\frac{1}{|r|^\varepsilon}\right)\,,
\end{eqnarray}
the boundary terms can be written as:
\begin{eqnarray}
 &&\frac{1}{8\pi}\int_{t_1}^{t_2}dt \int_0^\pi d\theta \int_0^{2\pi} d\varphi \left[ \widetilde{N}_+(t) 2 \widetilde{M}_+(t) \sin \theta + \widetilde{N}_-(t) 2 \widetilde{M}_-(t) \sin \theta\right]\nonumber\\
 &&= \int_{t_1}^{t_2}dt \left[\widetilde{N}_+(t) \widetilde{M}_+(t) +\widetilde{N}_-(t) \widetilde{M}_-(t)\right]\,.
\end{eqnarray}
For the standard Schwarzschild solution $\widetilde{M}_{\pm}$ coincides with the mass of the in standard static coordinates.

The canonical momenta are \cite{dirac1966,DeWitt1967,Esposito1992}:
\begin{eqnarray}
\widetilde{\pi}_{\Lambda} &\equiv&\frac{\partial {\mathcal{L}}_{EF}}{\partial \dot{\widetilde{\Lambda}}  } = -\frac{\widetilde{R} \big( \dot{\widetilde{R}} - \widetilde{R}' \widetilde{N}^{r} \big)}{\widetilde{N}}  \,,
\label{eq:PLambda:0} \\
\widetilde{\pi} _{R}&\equiv&\frac{\partial {\mathcal{L}}_{EF}}{\partial \dot{\widetilde{R}}  } = -\frac{\widetilde{\Lambda} \big(
\dot{\widetilde{R}} - \widetilde{R}' \widetilde{N}^{r} \big) + \widetilde{R} \big( \dot{\widetilde{\Lambda}} - (\widetilde{\Lambda} \widetilde{N}^{r})'
\big)}{\widetilde{N}}  \,,   
\label{eq:PiR:0} \\
\widetilde{\pi}_{\phi} &\equiv&\frac{\partial {\mathcal{L}}_{EF}}{\partial \dot{\widetilde{\phi}}  }   
= \frac{\widetilde{\Lambda} \widetilde{R}^{2}}{2 \widetilde{N}} 
\big( \dot{\widetilde{\phi}} - \widetilde{N}^{r} \widetilde{\phi}' \big) \,.
\label{eq:PPhi:0}
\end{eqnarray}
Following the standard procedure explained before, the Hamiltonian density constraint \eqref{H:const:0} becomes:
\begin{eqnarray}
\mathcal{\widetilde{H}}&=&-\frac{\widetilde{\pi}_{R} \widetilde{\pi}_{\Lambda}}{\widetilde{R}}  
+\frac{\widetilde{\Lambda} \widetilde{\pi}_{\Lambda}^{2}}{2 \widetilde{R}^{2}}
+\frac{\widetilde{R} \widetilde{R}''}{\widetilde{\Lambda}}  -\frac{\widetilde{R} \widetilde{R}' \widetilde{\Lambda}'}{\widetilde{\Lambda}^{2}} + \frac{\widetilde{R}'^{2}}{2\widetilde{\Lambda} } - \frac{\widetilde{\Lambda}}{2} 
+ \frac{\widetilde{\pi}^{2}_{\phi}}{\widetilde{\Lambda} \widetilde{R}^2}  
    +\frac{\widetilde{R}^{2}\widetilde{\phi}'^{2}}{4\widetilde{\Lambda}}\,,
\end{eqnarray}
and the momentum constraint \eqref{Hi:const:0} is:
\begin{eqnarray}
\mathcal{\widetilde{H}}_r=\widetilde{\pi}_{R} \widetilde{R}' - \widetilde{\Lambda} \widetilde{\pi}_{\Lambda}' + \widetilde{\pi}_{\phi}{\widetilde{\phi}}'\,.    
\end{eqnarray}

The equations of motion in EF are:
\begin{eqnarray}
\dot{\widetilde{\Lambda}} &\approx& -\frac{\widetilde{N} \widetilde{\pi}_R}{\widetilde{R}}+\frac{\widetilde{N} \widetilde{\Lambda} \widetilde{\pi}_\Lambda}{\widetilde{R}^2}+\left(\widetilde{N}^r \widetilde{\Lambda} \right)'\,,\label{eq_dotLambda_EF}\\
\dot{\widetilde{\pi}}_\Lambda  &\approx& -\frac{\widetilde{N} \widetilde{\pi}_\Lambda^2}{2 \widetilde{R}^2} + \frac{\widetilde{N} \widetilde{R} \widetilde{R''}}{\widetilde{\Lambda}^2} -2 \frac{\widetilde{N} \widetilde{R} \widetilde{R'} \widetilde{\Lambda'}}{\widetilde{\Lambda}^3}
-\left(\frac{\widetilde{N} \widetilde{R} \widetilde{R}'}{\widetilde{\Lambda}^2}\right)' +\frac{\widetilde{N} \widetilde{R'}^2}{2\widetilde{\Lambda}^2} +\frac{\widetilde{N}}{2}
+\frac{\widetilde{N} \widetilde{\pi}_\phi^2}{\widetilde{\Lambda}^2 \widetilde{R}^2} +\frac{\widetilde{N} \widetilde{R}^2 \widetilde{\phi'}^2}{4\widetilde{\Lambda}^2} + \widetilde{N}^r \widetilde{\pi'}_\Lambda\\
\dot{\widetilde{R}} &\approx&  -\frac{\widetilde{N} \widetilde{\pi}_\Lambda}{\widetilde{R}}+\widetilde{N}^r \widetilde{R'}\,,\\
\dot{\widetilde{\pi}}_R &\approx& -\frac{\widetilde{N} \widetilde{\pi}_\Lambda \widetilde{\pi}_R }{\widetilde{R}^2}+ \frac{\widetilde{N} \widetilde{\Lambda} \widetilde{\pi}_\Lambda^2}{R^3} - \frac{\widetilde{N} \widetilde{R''}}{\widetilde{\Lambda}}-\left(\frac{\widetilde{N} \widetilde{R}}{\widetilde{\Lambda}}\right)''
+\frac{\widetilde{N} \widetilde{R'} \widetilde{\Lambda'}}{\widetilde{\Lambda}^2} -\left( \frac{\widetilde{N} \widetilde{R} \widetilde{\Lambda'}}{\widetilde{\Lambda}^2}\right)'+\left(\frac{\widetilde{N} \widetilde{R'}}{\widetilde{\Lambda}} \right)'\nonumber\\
&& +2\frac{\widetilde{N} \widetilde{\pi}_\phi^2}{\widetilde{\Lambda} \widetilde{R}^3} -\frac{\widetilde{N} \widetilde{R} {\widetilde{\phi'}}^2}{2 \widetilde{\Lambda}} + \left(\widetilde{N}^r \widetilde{\pi}_R \right)'\,,\\
\dot{\widetilde{\phi}}&\approx& 2\frac{\widetilde{N} \widetilde{\pi}_\phi}{\widetilde{\Lambda} \widetilde{R}^2} + \widetilde{N}^r \widetilde{\phi'}\,,\\
\dot{\widetilde{\pi}}_\phi&\approx& \left(\frac{\widetilde{N} \widetilde{R}^2 \widetilde{\phi'}}{2 \widetilde{\Lambda}}\right)' + \left(\widetilde{N}^r \widetilde{\pi}_\phi \right)'\,.\label{eq_dotPiPhi_EF}
\end{eqnarray}

The equations of motion for $\widetilde{\Lambda}$, $\widetilde{\pi}_\Lambda$, $\widetilde{R}$ and $\widetilde{\pi}_R$  - when there is no scalar field (General Relativity) - coincide with the equations derived in \cite{Loll:2017utf} (modulo a constant).

\section{ADM analysis in the Jordan frame}
\label{section:JF}

In the Jordan frame, our starting point is the following scalar-tensor Lagrangian \cite{Callan:1970ze,Bekenstein:1974sf,Bronnikov:2002kf,Faraoni:2004pi,Galtsov:2020jnu,Ray:2024fyx}:  
\begin{eqnarray}
\label{Lagr:JF}
\mathcal{L}_{JF}=\frac{1}{16\pi} \left[\left(1-\frac{\phi^2}{6}\right){}^{(4)}R-g^{\mu\nu}\partial_\mu\phi\partial_\nu\phi \right]\,.
\end{eqnarray}
In this theory the equation of motion for the scalar field $\phi$ is invariant under conformal transformations \cite{Callan:1970ze,Bekenstein:1974sf}.
This model  has received attention in the literature because it is the only theory of gravity, and all other metric theories,  for which the equivalence principle holds when $\phi$ is not a gravitational field \cite{Faraoni:2004pi}.
Moreover, this scalar-tensor theory is, in general, required by the renormalization of the theory \cite{Faraoni:2004pi}.

The action functional with boundary terms which make it linear in the variation of the metric tensor in this case is:
\begin{eqnarray}
S_{JF} &=&\frac{1}{16\pi }
\int_{{M}}d^{4}x \sqrt{-g}\; \left[\left(1-\frac{\phi^2}{6}\right){}^{(4)}R-g^{\mu\nu}\partial_\mu\phi\partial_\nu\phi \right]
+\frac{1}{8\pi  }\int_{\Sigma_{t_1}}d^{3}x\sqrt{h}\left(1-\frac{\phi^2}{6}\right)K \nonumber\\
&&-\frac{1}{8\pi }\int_{\Sigma_{t_0}}d^{3}x\sqrt{h}\left(1-\frac{\phi^2}{6}\right)K - \frac{1}{8\pi  }\int_{B}d^{3}x\sqrt{-\gamma}\,\left(1-\frac{\phi^2}{6}\right)\Theta \nonumber\\
&&- \frac{1}{8\pi }\int_{B_{t_0}} d^2x \sqrt{\sigma} \left(1-\frac{\phi^2}{6}\right)\, \mathrm{arcsinh} (n^\mu u_\mu)  
+\frac{1}{8\pi }\int_{B_{t_1}} d^2x \sqrt{\sigma} \left(1-\frac{\phi^2}{6}\right) \,\mathrm{arcsinh} (n^\mu u_\mu).
 \label{JF:tutta}
\end{eqnarray}
We consider here the following $3+1$ decomposition of the trace of the Ricci tensor \cite{DeWitt1967}:
\begin{equation}
\sqrt{-g}{}^{(4)}R=\sqrt{h}N\left({}^{3}R+K_{ij}K^{ij}-K^{2}\right)-(2K\sqrt{h}),_{0}+2f^{i},_{i} \,,
\label{ariscompongo} 
\end{equation}
where $f^{i}$ is defined as: 
\begin{equation}
f^{i}\equiv \sqrt{h}(KN^{i}-h^{ij}N,_{j})\,.
\label{def:fi}
\end{equation}
This decomposition is analogous to the one defined in Eq. \eqref{covo}, see Appendix~\ref{App:1}. 

The bulk term of the previous action can be rewritten as:
\begin{eqnarray}
&&\frac{1}{16\pi}\int dt \int_{\Sigma}N\sqrt{h}d^{3}x 
    \Bigg[\left(1-\frac{\phi^2}{6}\right)\left({}^{(3)}R+K_{ij}K^{ij}-K^{2}\right)
    -g^{00}\partial_{0} \phi\partial_{0} \phi-2g^{0r}\partial_{0}\phi\partial_{r}\phi
    \nonumber\\    &&
    -h^{rr}\partial_{r} \phi\partial_{r} \phi\Bigg]
    +\frac{1}{16\pi}\int dt \int_{\Sigma}d^{3}x\left(1-\frac{{\phi}^{2}}{6}\right)\left(-(2K\sqrt{h}),_{0}+2f^{i},_{i} \right)\\
  &=&\frac{1}{16\pi}\int dt \int_{\Sigma}N\sqrt{h}d^{3}x 
    \Bigg[\left(1-\frac{\phi^2}{6}\right)\left({}^{(3)}R+K_{ij}K^{ij}-K^{2}\right)  
  -g^{00}\partial_{0} \phi\partial_{0} \phi-2g^{0r}\partial_{0}\phi\partial_{r}\phi\nonumber\\
  && -h^{rr}\partial_{r} \phi\partial_{r} \phi\Bigg]
-\frac{1}{8\pi}\int dt \int_{\Sigma}d^{3}x\left\{\nabla_{0}\left[\left(1-\frac{{\phi}^{2}}{6}\right)K\sqrt{h}\right] \right\}\nonumber\\
&&+\frac{1}{8\pi}\int dt \int_{\Sigma}d^{3}x\left\{\nabla_{0}\left[\left(1-\frac{{\phi}^{2}}{6}\right)\right]K\sqrt{h} \right\}
+\frac{1}{8\pi}\int dt \int_{\Sigma}d^{3}x\left\{\nabla_{i}\left[\left(1-\frac{{\phi}^{2}}{6}\right)f^{i}\right] \right\}\nonumber\\
    &&-\frac{1}{8\pi}\int dt \int_{\Sigma}d^{3}x\left\{\nabla_{i}\left[\left(1-\frac{{\phi}^{2}}{6}\right)\right]f^{i} \right\}\,.
\end{eqnarray}
In this JF case it is crucial to consider properly all the integration by parts terms, or else some contributions are overlooked. 

The bulk term of the physical action \eqref{phys:action:1} for the spherically symmetric ADM metric introduced in Eq.~\eqref{def:ADMss}, integrating over $\theta$ and $\varphi$, becomes:
\begin{eqnarray}
S^{\mathrm{PHYS}}_{JF}
&=& \int dt \int _{-
\infty} ^{\infty} dr \Bigg\{ 
\left(1-\frac{\phi ^2}{6}\right) \left[-\frac{1}{N}\left(\frac{\Lambda}{2}\left(-\dot{R}+R^\prime N^r\right)^2+R\left(-\dot{\Lambda}+(\Lambda N^r)^\prime\right)\left(-\dot{R}+R^\prime N^r\right)\right)\right.\nonumber\\
&&\left. +N \left(-\frac{R R^{\prime}}{\Lambda}+\frac{R R^\prime \Lambda^\prime}{\Lambda^2}-\frac{R^{\prime 2}}{2\Lambda}+\frac{\Lambda}{2} \right) \right] +\frac{1}{4}\left(-\frac{N R^2 \phi^{\prime 2}}{\Lambda}+\frac{\Lambda R^2}{N}(\dot{\phi}-N^r \phi^\prime)^2\right)\nonumber\\
&&-\frac{1}{6}\phi \frac{ R}{N}\left[R\left(-\dot{\Lambda}+(\Lambda N^r)^\prime\right)+2\Lambda\left(-\dot{R}+R^\prime N^r\right)\right] \left(\dot{\phi}-N^r\phi^\prime\right)\nonumber\\
&&+\frac{1}{6}N\Lambda R^2 \left(\frac{\phi^{\prime 2}}{\Lambda^2}+\frac{\phi \phi^{\prime \prime}}{\Lambda^2}-\frac{\Lambda^\prime \phi \phi^\prime}{\Lambda^3}+\frac{2 \phi R^\prime \phi^\prime}{R \Lambda^2}\right)
\Bigg\}\,. 
\label{eq:S:JF1}
\end{eqnarray}
Taking into account all the terms present in the previous Eq.~\eqref{eq:S:JF1}, we derive the canonical momenta associated to this action are:
\begin{eqnarray}
\pi_\Lambda&\equiv&\frac{\partial \mathcal{L}_{JF}}{\partial \dot{\Lambda}}
=-\frac{1}{N}\left(1-\frac{\phi ^2}{6}\right) R\left(\dot{R}-R^{\prime} N^r\right)+\frac{R^2 \phi}{6 N} (-N^r \phi^\prime+\dot{\phi})\,, \label{eq:PLambda:JF} \\
\pi_R&\equiv&\frac{\partial \mathcal{L}_{JF}}{\partial \dot{R}}=
\frac{1}{N}\left(1-\frac{\phi ^2}{6}\right)\left[R\left(-\dot{\Lambda}+(\Lambda N^r)^\prime\right)+\Lambda\left(-\dot{R}+R^\prime N^r\right)\right]\nonumber\\
&&\qquad\qquad +\frac{\phi \Lambda R}{3 N}\left(\dot{\phi}-N^r \phi^\prime\right)\,,\\
\pi_\phi&\equiv&\frac{\partial \mathcal{L}_{JF}}{\partial \dot{\phi}}=
-\frac{\phi R}{6 N}\left[R\left(-\dot{\Lambda}+(\Lambda N^r)^\prime\right)+2\Lambda\left(-\dot{R}+R^\prime N^r\right)\right]\nonumber\\
&&\qquad\qquad +\frac{\Lambda R^2}{2 N}\left(\dot{\phi}-N^r \phi^\prime\right)\,.
\label{eq:PPhi:JF}
\end{eqnarray}
In previous literature some terms due to the integration by parts of the action were neglected,
therefore some terms were missing in the definition of momenta \cite{Esposito:2007xz}.
These integration by parts contributions do not affect the the EF the result, but the effects are evident in this JF case.
If those terms are missing in the definition of momenta it is not possible to find the usual relation among canonical momenta in the two frames, see Eqs.~\eqref{EFJF:momenta1}-\eqref{EFJF:momenta2}  below and also \cite{Gionti2021,Galaverni:2021xhd,GiontiSJ:2023tgx} for the Brans-Dicke theory case.

Following the procedure discussed in Section \ref{section:2} we obtain the expression for the  Hamiltonian density constraint \eqref{H:const:0} in this particular case:
\begin{eqnarray}
\mathcal{H}&=&\frac{\phi^2\pi_R^2}{36\Lambda}\left(1-\frac{\phi ^2}{6}\right)^{-1}+\frac{\Lambda\pi^2_\Lambda}{2R^2} \left(1+\frac{\phi^2}{18}\right)\left(1-\frac{\phi^2}{6}\right)^{-1}+\frac{\pi_\phi^2}{R^2\Lambda} \left(1-\frac{\phi ^2}{6}\right)
\nonumber\\
&&-\frac{\pi_R\pi_\Lambda}{R}\left(1-\frac{\phi^2}{18}\right)\left(1-\frac{\phi^2}{6}\right)^{-1}+\frac{\phi\pi_R \pi_\phi}{3 R \Lambda}+\frac{\phi \pi_\Lambda\pi_\phi}{3 R^2}\\
&&+\left(1-\frac{\phi ^2}{6}\right)\left(-\frac{\Lambda}{2}+\frac{R^{\prime 2} }{2\Lambda}-\frac{R R^\prime \Lambda^\prime}{\Lambda^2}+\frac{R R^{\prime\prime}}{\Lambda}\right)+\frac{R^2\phi^{\prime 2}}{12 \Lambda}
+\frac{R^2 \phi \Lambda^\prime \phi^\prime}{6\Lambda^2}-\frac{R^2 \phi \phi^{\prime\prime}}{6\Lambda}
-\frac{R\phi R^\prime \phi^\prime}{3\Lambda}\,,\nonumber
\end{eqnarray}
and the momentum constraint \eqref{Hi:const:0}:
\begin{eqnarray}
\mathcal{H}_r=\pi_{R} R' - \Lambda \pi_{\Lambda}' + \pi_{\phi}{\phi}'\,.    
\end{eqnarray}

The conformal transformation connecting EF and JF cases is the following \cite{Bekenstein:1974sf,Galtsov:2020jnu}:
\begin{equation}
\widetilde{g}_{\mu\nu}=\left(1-\frac{{\phi}^2}{6}\right)g_{\mu\nu}\,,\,\mathrm{and}\,\,
\widetilde{\phi}=\sqrt{6} \tanh^{-1}\frac{\phi}{\sqrt{6}}\,,
\label{def:conf:mphi}
\end{equation}
corresponding to the following relations:
\begin{eqnarray}
&&\widetilde{N}=\left(1-\frac{{\phi}^2}{6}\right)^{\frac{1}{2}}N\,,\,\,{\widetilde{N}}^r=N^r\,,\,\,\nonumber\\
&&\widetilde{\Lambda}=\left(1-\frac{{\phi}^2}{6}\right)^{\frac{1}{2}}\Lambda\,,\,\,\widetilde{R}=\left(1-\frac{{\phi}^2}{6}\right)^{\frac{1}{2}}R\,.    
\label{def:conf:metric}
\end{eqnarray}
Comparing the definitions of canonical momenta in the EF Eqs.~\eqref{eq:PLambda:0}-\eqref{eq:PPhi:0} and in the JF Eqs.~\eqref{eq:PLambda:JF}-\eqref{eq:PPhi:JF} we get this relations among canonical momenta in the two frames \cite{Gionti2021,Galaverni:2021xhd,GiontiSJ:2023tgx}:
\begin{eqnarray}
\label{EFJF:momenta1}
{\widetilde{\pi}}_{\Lambda}&=&\left(1-\frac{{\phi}^2}{6}\right)^{-\frac{1}{2}}{\pi}_{\Lambda}\,,\,\,\\
{\widetilde{\pi}}_R&=&\left(1-\frac{{\phi}^2}{6}\right)^{-\frac{1}{2}}\pi_R\,,\\
\widetilde{\pi}_{\phi}&=& \left( 1-\frac{{\phi}^2}{6} \right)\pi_{\phi}+\frac{1}{6}R\,\phi\,\pi_{R}+\frac{1}{6}\Lambda\,\phi\,\pi_{\Lambda}\,. \label{EFJF:momenta2}   
\end{eqnarray}

If we calculate that the Poisson brackets of the EF canonical variables expressed as function of the variables in the JF we easily see that:
\begin{eqnarray}
\left\{\widetilde{N},{\widetilde{\pi}}_{\phi} \right\}=-N\left(1-\frac{\phi^{2}}{6}\right)^{\frac{1}{2}} \frac{\phi}{6}\,,\\
\left\{\widetilde{N}_r,{\widetilde{\pi}}_{\phi} \right\}=-N_r \left(1-\frac{\phi^{2}}{6}\right)\frac{\phi}{3}\,.
\end{eqnarray}
Therefore, the Hamiltonian canonical transformations from the Jordan to the Einstein frame cannot be considered a canonical transformation strictly speaking, see detailed discussion in \cite{Gionti2021,Galaverni:2021xhd,GiontiSJ:2023tgx}.

We can avoid this problem considering a reduced phase space where we gauge fix the lapse function $N$ and the shift function $N^r$, implementing them as secondary constraints $\chi_i$ as discussed in \cite{GiontiSJ:2023tgx}.
Equations of motion are derived using Dirac brackets (DB), defined starting from the Poisson brackets and the inverse of the second class constraint matrix ($C_{\alpha\beta}\equiv\{\chi_\alpha,\chi_\beta\}$) \cite{dirac1966,Henneaux:1992ig,GiontiSJ:2023tgx}:
\begin{equation}
\{\cdot,\cdot\}_{DB}\equiv \{\cdot,\cdot\}- \{\cdot,\chi_\alpha \} C_{\alpha\beta}^{-1} 
\{\chi_\beta,\cdot\}\,.
\end{equation}
 Afterward - strongly imposing the second class constraints - we reduce the dynamical variables to: $\{ \Lambda,R,\phi,\pi_\Lambda,\pi_R,\pi_\phi \}$. On this reduced phase space, defined by gauge fixing the lapse and the shift function, the transformation from EF and JF is Hamiltonian canonical.

Therefore equations of motion in the JF are:
\begin{eqnarray}
\dot{\Lambda} &\approx& \frac{N\Lambda \pi_\Lambda}{R^2}\left(1+\frac{\phi^2}{18}\right)\left(1-\frac{\phi^2}{6}\right)^{-1}
-\frac{N \pi_R}{R}\left(1-\frac{\phi^2}{18}\right)\left(1-\frac{\phi^2}{6}\right)^{-1}
+\frac{N\phi\pi_\phi}{3R^2}
-\left(N^r \Lambda\right)' \,,\label{eq_dotLambda_JF}\\
\dot{\pi}_\Lambda&\approx& 
\frac{N\phi^2 \pi_R^2}{36\Lambda^2} \left(1-\frac{\phi^2}{6}\right)^{-1}
-\frac{N\pi_\Lambda^2}{3 R^2} \left(1+\frac{\phi^2}{18}\right)\left(1-\frac{\phi^2}{6}\right)^{-1}
+\frac{N \pi_\phi^2}{R^2 \Lambda^2}\left(1-\frac{\phi^2}{6}\right)\nonumber\\
&& +\frac{N }{2}\left(1-\frac{\phi^2}{6}\right)
+\frac{N R'^2}{2 \Lambda^2}\left(1-\frac{\phi^2}{6}\right)
-\frac{2 N R R' \Lambda'}{\Lambda^3}\left(1-\frac{\phi^2}{6}\right)\nonumber\\
&&-\left[\frac{N R R'}{\Lambda^2}\left(1-\frac{\phi^2}{6}\right) \right]'
+\frac{N R R''}{\Lambda^2}\left(1-\frac{\phi^2}{6}\right)
+\frac{N R^2 \phi'^2}{12 \Lambda^2}+\frac{N R^2 \phi \Lambda' \phi'}{3 \Lambda^3}
-\frac{N R \phi R' \phi'}{3 \Lambda^2}\nonumber\\
&&+\left(\frac{N R^2 \phi \phi'}{6 \Lambda^2}\right)'
-\frac{N R^2 \phi \phi''}{6 \Lambda^2}
+\frac{N\phi \pi_R \pi_\phi}{3 R \Lambda^2}
+N^r \pi'_\Lambda\,,\\
\dot{R}&\approx&
\frac{N\pi_R \phi^2}{18 \Lambda}   \left(1-\frac{\phi^2}{6}\right)^{-1}
-\frac{N\pi_\Lambda}{R} \left(1-\frac{\phi^2}{18}\right)\left(1-\frac{\phi^2}{6}\right)^{-1}
+\frac{N\phi\pi_\phi}{3 R \Lambda}
+N^r R'\,,\\
\dot{\pi}_R&\approx&
\frac{N\Lambda \pi_\Lambda^2}{ R^3 }\left(1+\frac{\phi^2}{18}\right)\left(1-\frac{\phi^2}{6}\right)^{-1}
+\frac{2 N \pi_\phi^2}{R^3 \Lambda} \left(1-\frac{\phi^2}{6}\right)
-\frac{N\pi_R \pi_\Lambda}{R^2} \left(1-\frac{\phi^2}{18}\right)\left(1-\frac{\phi^2}{6}\right)^{-1}\nonumber\\
&&+\frac{N\phi \pi_R \pi_\phi }{3 R^2 \Lambda}
+\frac{2 N \phi \pi_\Lambda \pi_\phi}{3 R^3 }
+\left[\frac{N R'}{\Lambda} \left(1-\frac{\phi^2}{6}\right)  \right]'
+\frac{N R' \Lambda'}{\Lambda^2} \left(1-\frac{\phi^2}{6}\right)
-\left[\frac{N R \Lambda'}{\Lambda^2}\left(1-\frac{\phi^2}{6}\right)\right]' \nonumber\\
&&-\frac{N R''}{\Lambda}\left(1-\frac{\phi^2}{6}\right)
-\left[\frac{N R}{\Lambda}\left(1-\frac{\phi^2}{6}\right)\right]''
-\frac{N R \phi'^2}{6 \Lambda}
-\frac{N R \phi \Lambda' \phi'}{3 R^2}
+\frac{N R \phi \phi''}{3 \Lambda}
+\frac{N\phi R' \phi'}{3\Lambda}\nonumber\\
&&-\left(\frac{N R \phi \phi'}{3\Lambda}\right)'
+\left(N^r \pi_R\right)'\,,\\
\dot{\phi}&\approx&
\frac{2 N \pi_\phi}{R^2 \Lambda} \left(1-\frac{\phi^2}{6}\right)
+\frac{N \pi_R \phi}{3 R \Lambda}
+\frac{N\phi \pi_\Lambda}{3 R^2}
+N^r \phi'\,,\\
\dot{\pi}_\phi&\approx& 
-\frac{N\pi_R^2 \phi}{18 \Lambda}  \left(1-\frac{\phi^2}{6}\right)^{-2}
-\frac{2 N \Lambda \phi \pi_\Lambda^2}{9 R^2}  \left(1-\frac{\phi^2}{6}\right)^{-2}
+\frac{N\phi \pi_\phi^2}{3 R^2 \Lambda}
+\frac{2 N \phi \pi_R \pi_\Lambda }{9 R}  \left(1-\frac{\phi^2}{6}\right)^{-2}\nonumber\\
&& -\frac{N \pi_R \pi_\phi}{3 R \Lambda}
-\frac{N \pi_\Lambda \pi_\phi}{3 R^2}
+\frac{N\phi}{3}\left(-\frac{\Lambda}{2}+ \frac{{R'}^2}{2 \Lambda}-\frac{R R' \Lambda'}{\Lambda^2}+\frac{R R''}{\Lambda}\right)
+\left(\frac{N R^2\phi'}{6\Lambda}\right)'
-\frac{N R^2 \Lambda' \phi'}{6\Lambda^2}\nonumber\\
&&+\left(\frac{N R^2 \phi \Lambda'}{6\Lambda^2}\right)'
+\frac{N R^2 \phi''}{6 \Lambda}
+\left(\frac{NR^2 \phi}{6 \Lambda}\right)''
+\frac{N R R' \phi'}{3 \Lambda}
-\left(\frac{N R \phi R'}{3 \Lambda}\right)'
+\left(N^r \pi_\phi\right)'\,.\label{eq_dotPiPhi_JF}
\end{eqnarray}

\end{widetext} 

\section{Static solutions and singular transformation}
\label{Sect:sol}

We see  that the equations of motion in the EF are verified by a particular static solution. 
Using the conformal transformation between the two frames a static solution in the JF obtained.
Finally, we show that the solutions in the two frames have a different structure (number of singularity); 
this is due to the fact that the Weyl (conformal) transformation is not well defined for a particular value of the radial coordinate.

\subsection{A static solution in the EF: FJNW solution}

A particular static solution of the Einstein General Relativity plus a massless scalar field, see the EF Lagrangian defined in 
Eq.~\eqref{EF:Lagr}, was studied by Fisher \cite{Fisher:1948yn} and Janis, Newman and
Winicour \cite{Janis:1968zz,Wyman:1981bd,Virbhadra:1997ie} and is is known as FJNW solution:
\begin{eqnarray}
\label{sol:FJNW:EF1}
d\widetilde{s}^2 &=& -\left(1-\frac{b}{r}\right)^{\gamma} dt^2
                     +\left(1-\frac{b}{r}\right)^{-\gamma} dr^2\nonumber\\
                     &&+r^2 \left(1-\frac{b}{r}\right)^{1-\gamma} d\Omega^2\,,\\
\widetilde{\phi}&=&\sqrt{\frac{1-\gamma^2}{2}}\ln\left(1-\frac{b}{r}\right)\,.
\label{sol:FJNW:EF2}
\end{eqnarray}
which, in general, is defined for $r<0$ and $r>b$.
The constants $b$ and $\gamma$ are related to the mass of the compact gravitational object ($m$) and
the scalar charge ($q$) \cite{Bekenstein:1975ts} by the following expressions \cite{Nandi:2006ds}:
\begin{equation}
b=2\sqrt{m^2+\frac{q^2}{2}}\,,\,\,\gamma=\frac{2m}{b}\,,    
\end{equation}
it is also useful to remember the following relation \cite{Galtsov:2020jnu}: $\gamma=\left(1-\frac{2q^2}{b^2}\right)^{1/2}$,
in general for $q\neq 0$ and real we have $0<\gamma<1$, and the FJNW metric is valid for $r>b$, provided the definition of the spherical radius \eqref{sol:FJNW:Lambda:EF1}  \cite{Pal:2022cxb}.
Note that for $q=0$, corresponding to $\gamma=1$, the FJNW solution \eqref{sol:FJNW:EF1} reduces to the Schwarzschild solution.

It is easy to verify that the equations of motion in the EF, Eqs.~\eqref{eq_dotLambda_EF}-\eqref{eq_dotPiPhi_EF}, are verified by this FJNW solution static solution if we replace:
\begin{eqnarray}
\label{sol:FJNW:N:EF1}
\widetilde{N}&=&\left(1-\frac{b}{r}\right)^{\gamma/2}\,,\;\; \widetilde{N}^r=0\,,\\
\label{sol:FJNW:Lambda:EF1}
\widetilde{\Lambda}&=& \left(1-\frac{b}{r}\right)^{-\gamma/2}\,,\;\; \widetilde{R}=r\left(1-\frac{b}{r}\right)^{(1-\gamma)/2}\,,\\
\widetilde{\phi}&=& \sqrt{\frac{1-\gamma^2}{2}}\ln \left(1-\frac{b}{r} \right)\,.
\label{sol:FJNW:phi:EF1}
\end{eqnarray}
Note that all momenta vanish since FJNW is a static solution.

\subsection{A static solution in the JF: BBMB solution}

Starting from the FJNW solution of the EF case, see Eqs.~\eqref{sol:FJNW:N:EF1}-\eqref{sol:FJNW:phi:EF1}, we obtain a solution in the JF.
Using relation between EF and JF introduced in Eq.~\eqref{def:conf:mphi},
we derive the following expression for the scalar field in the JF:
\begin{eqnarray}
\phi&=&\sqrt{6} \tanh\left[\frac{1}{\sqrt{6}}
\sqrt{\frac{1-\gamma^2}{2}}
\ln\left(1-\frac{b}{r} \right)\right]\nonumber\\
&=&\sqrt{6}\left[
1-\frac{2}{1+\left(1-\frac{b}{r}\right)^{\sqrt{\frac{1-\gamma^2}{3}}}}
\right]\,,
\label{phigamma:JF}
\end{eqnarray}
while the other canonical variables in the JF are obtained using Eqs.~\eqref{def:conf:metric}:
\begin{eqnarray}
N(r)&=&\frac{\left(1-\frac{b}{r}\right)^{\frac{\gamma}{2}}}{\sqrt{1-\left( 1-\frac{2}{1+\left(1-\frac{b}{r}\right)^{\sqrt{\frac{1-\gamma^2}{3}}}} \right)^2}} \,,\;\;\\ N^r(r)&=&0\,,\\
\Lambda(r)&=&\frac{\left(1-\frac{b}{r}\right)^{-\frac{\gamma}{2}}}{\sqrt{1-\left( 1-\frac{2}{1+\left(1-\frac{b}{r}\right)^{\sqrt{\frac{1-\gamma^2}{3}}}} \right)^2}}\,, \;\;\\
R(r)&=&\frac{r\left(1-\frac{b}{r}\right)^{\frac{1-\gamma}{2}}}{\sqrt{1-\left( 1-\frac{2}{1+\left(1-\frac{b}{r}\right)^{\sqrt{\frac{1-\gamma^2}{3}}}} \right)^2}} \,.
\label{Rgamma:JF}
\end{eqnarray}
It is easy to check that \eqref{phigamma:JF}-\eqref{Rgamma:JF} verify the equations of motion ($\mathcal{H}=0$) in the JF case: \eqref{eq_dotLambda_JF}-\eqref{eq_dotPiPhi_JF}.
 
\begin{figure}[htb]
    \centering
    \includegraphics[width=0.9\linewidth]{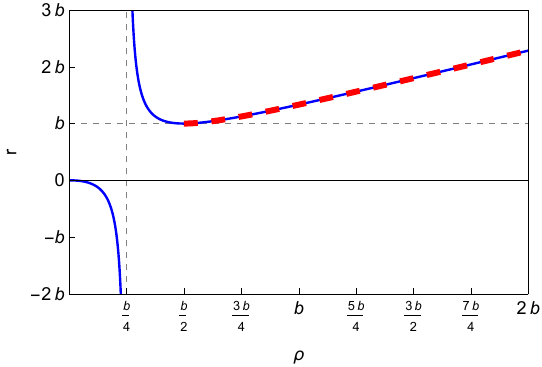}
    \caption{Coordinate $r$ as a function of the new radial coordinate $\rho$ defined in Eq. \eqref{def:rho}.
    The conformal transformation  \eqref{def:conf:mphi}-\eqref{def:conf:metric} between JF and EF is well posed for $\rho>b/2$, therefore the FJNW manifold $r>b$ is mapped to the part of BBMB manifold with $\rho>b/2$ (red-dashed line).
    }
    \label{fig:rho}
\end{figure}

For the particular case $\gamma=1/2$ 
\cite{Galtsov:2020jnu}
the Eq.~\eqref{phigamma:JF} for the scalar field becomes:
\begin{equation}
\phi=\sqrt{6} \tanh\left[\frac{1}{4}\ln\left(1-\frac{b}{r} \right)\right]=-\sqrt{6}\frac{1-\sqrt{1-b/r}}{1+\sqrt{1-b/r}}\,,    
\end{equation}
while the other canonical variables in the JF are:
\begin{eqnarray}
&&N(r)=\frac{1}{2}\left(1+\sqrt{1-\frac{b}{r} }\right) \,,\;\; N^r(r)=0\,,\\
&&\Lambda(r)=\frac{1}{2}\left(1+\frac{1}{\sqrt{1-\frac{b}{r} }}\right)\,, \;\;\\
&&R(r)=\frac{r}{2}\left(1+\sqrt{1-\frac{b}{r} }\right) \,.
\end{eqnarray}
In this $\gamma=1/2$ case it is useful to introduce a new coordinate $\rho$ defined as:
\begin{equation}
\label{def:rho}
 \sqrt{1-\frac{b}{r}} \equiv   1-\frac{b}{2 \rho}\,.     
\end{equation}
In terms of this new coordinate $\rho$ the canonical variables in the JF are:
\begin{eqnarray}
&&N(\rho)=1-\frac{b}{4\rho}\,,\;\; N^r(\rho)=0\,,\\
&&\Lambda(\rho)=\frac{1}{2}\left(1+\frac{1}{\sqrt{1-\frac{b}{r} }}\right)\frac{dr}{d\rho}= \left(1-\frac{b}{4\rho}\right)^{-1}\,,\\
&& R(\rho)=\rho\,,\;\;\phi(\rho)=-\sqrt{6}\frac{ b/4}{\rho-b/4} \,,
\end{eqnarray}
and the solution takes the BBMB (Bocharova-Bronnikov-Melnikov-Bekenstein) form \cite{Bocharova:1970,Bekenstein:1974sf,Galtsov:2020jnu}: 
\begin{eqnarray}
ds^2&=&-\left(1-\frac{b}{4\rho}\right)^2 dt^2
       +\left(1-\frac{b}{4\rho}\right)^{-2} d\rho^2\nonumber\\
       &&+\rho^2 d\Omega^2\,, \label{BBMB1}\\
\phi&=&-\sqrt{6}\frac{\frac{b}{4}}{\rho-\frac{b}{4}}\,.\label{BBMB2}
\end{eqnarray}
In this case $ R(\rho)=\rho>0$:
$\rho=b/4$ corresponds to the event horizon of the BBMB solution \cite{Turimov:2025odi}. 
For this particular value of the radial coordinate, the relation between $\rho$ and $r$ is singular, see Eq. \eqref{def:rho} 
and Fig. \ref{fig:rho}:
\begin{equation}
\frac{r}{b}=\left[1-\left(1-\frac{b}{2\rho}\right)^2\right]^{-1}\,.    
\end{equation}

\subsection{Peculiar aspects of the transformation between JF and EF}
\label{SubSect:sing}

We study here the properties of the static solutions presented earlier and evaluate the curvature invariants.
Starting from the Ricci tensor ($R_{\mu\nu}$) and the Riemann tensor ($R_{\mu\nu\rho\sigma}$) we obtain the Ricci scalar, the Ricci square and the Kretschmann scalar \cite{Turimov:2022rrx}.

For the FJNW solution in the EF, see Eq.~\eqref{sol:FJNW:EF1}, we get the following invariants:
\begin{eqnarray}
\widetilde{g}^{\mu\nu}\widetilde{R}_{\mu\nu}&=& \frac{1-\gamma^2}{2(b-r)^2} \left(1-\frac{b}{r}\right)^\gamma \left(\frac{b}{r}\right)^2 
\xrightarrow{\: \gamma = 1 \: } 0\,,\label{R1:FJNW}\\
\widetilde{R}^{\mu\nu}\widetilde{R}_{\mu\nu}&=& \frac{(1-\gamma^2)^2}{4(b-r)^4} \left(1-\frac{b}{r} \right)^{2 \gamma} \left(\frac{b}{r}\right)^4  
\xrightarrow{\: \gamma = 1 \: } 0\,,\\
\widetilde{R}^{\mu\nu\rho\sigma}\widetilde{R}_{\mu\nu\rho\sigma}&=& \left[12\gamma^2 + \frac{(1+\gamma)^2(3+2\gamma+7\gamma^2)}{4}\left(\frac{b}{r}\right)^2 \right.\nonumber\\
&&\left. -4\gamma (1+\gamma)(1+2\gamma)\frac{b}{r} \right]\nonumber\\
&&\frac{1}{(b-r)^4 } \left(1-\frac{b}{r}\right)^{2\gamma} \left(\frac{b}{r}\right)^2\nonumber\\
&& \xrightarrow{\: \gamma = 1 \: } \frac{12b^2}{r^6}=\frac{48 m^2}{r^6} \label{R3:FJNW}\,.
\end{eqnarray}

Note that these invariants have two singular points:
at $r=0$ - ouside the range of the solutions of FJNW metric - 
and also at $r=b$ (if $\gamma\neq 1$). 
This $r=b$ curvature singularity is a naked singularity since the metric does not have any event horizon \cite{Virbhadra:1997ie,Galtsov:2020jnu,Pal:2022cxb}.

If we evaluate the curvature invariant for the BBMB solution in the JF, see Eq.~\eqref{BBMB1}, we obtain: 
\begin{eqnarray}
g^{\mu\nu}R_{\mu\nu}&=& 0\,,\\
R^{\mu\nu}R_{\mu\nu}&=& \frac{b^4}{64 \rho^8} \,,\\
R^{\mu\nu\rho\sigma}R_{\mu\nu\rho\sigma}&=& \frac{b^2(7b^2-48 b \rho + 96 \rho^2)}{32 \rho^8}\,.
\end{eqnarray}
BBMB invariants appear to have only one curvature singularity at $\rho=0$.

\begin{figure}[htb]
    \centering
    \includegraphics[width=0.9\linewidth]{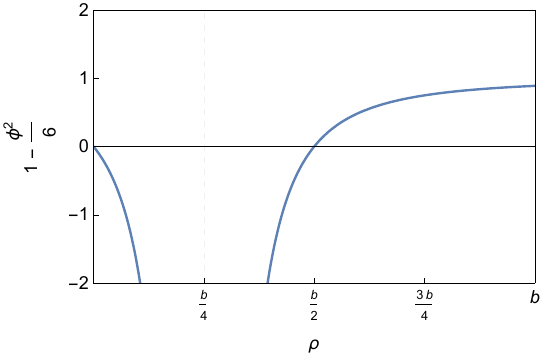}
    \caption{Conformal factor $(1-\frac{\phi^2}{6})$, see Eq. \eqref{def:conf:mphi}, as a function of the new radial coordinate $\rho$.}
    \label{fig:conformal}
\end{figure}

The conformal transformation \eqref{def:conf:mphi} connects two metrics with a different number of curvature singularities (different physical structure). 
This is due to the fact that for the FJNW naked singurality corresponding to $r=b$ (or to $\rho=\frac{b}{2}$ in the new radial coordinate) 
the conformal transformation \eqref{def:conf:mphi} connecting EF and JF is not well defined:
\begin{equation}
\phi(\rho)\xrightarrow{\: \rho = b/2 \: } -\sqrt{6}\,,
\mbox{therefore:}\,
\left(1-\frac{\phi^2}{6}\right)\xrightarrow{\: \rho = b/2 \: } 0\,.
\end{equation}
For $\rho=\frac{b}{2}$ ($r=b$) the conformal factor vanishes, see also Fig. \ref{fig:conformal}, and it is not possible to connect the two frames with the conformal transformation \eqref{def:conf:mphi}. 
Moreover, the scalar field $\widetilde{\phi}$ in the EF diverges, see Eqs.~\eqref{sol:FJNW:EF2}. 
 In other words, the conformal transformation between JF and EF \eqref{def:conf:metric} is well posed when  $(1-\frac{\phi^2}{6})>0$, or $\rho>b/2$ (see Fig. \ref{fig:conformal}). Therefore the whole FJNW manifold $r>b$ is mapped to a part of the BBMB manifold corresponding to $\rho>b/2$, see also Fig. \ref{fig:rho}.

As we have already remarked, in scalar tensor theories of gravity we avoid the value of the scalar field for which the Weyl(conformal) factor is null. Anyway, at the level of the solutions of the equations of motion, conformal continuation of solutions in these regions has to be carefully studied \cite{Bronnikov:2002kf}.

\section{Conclusions}
\label{Conclusions}

In this work, we have provided a short review of the ADM Hamiltonian formalism and applied it to spherically symmetric systems.
We discussed in detail the contribution of all the boundary terms in the most general case when space-like and time-like foliations are not orthogonal (see Fig. \ref{fig:1}). We also highlighted the importance of taking into account all the contributions coming from boundary terms and integration by parts. As we showed for the JF case in Section \ref{section:JF}, these terms are crucial in order to obtain the correct expression for canonical momenta and equations of motion. 

We derived the equations of motion for general non-static spherically symmetric systems both in EF and in JF  for a particular scalar-tensor theory where the scalar field is conformally coupled to gravity.
The algebra of the Poisson brackets in the JF among the Hamiltonian canonical variables in the EF 
is the same as in the general case without spherical symmetry. This is a check that the calculations have been performed correctly. 

The Weyl (conformal) Hamiltonian transformation connecting the EF and the JF is Hamiltonian canonical 
when a gauge fixing on the lapse $N$ and shift $N^r$ is performed.
These conditions are implemented as secondary constraints, and the relative Dirac’s Brackets substitute the Poisson's brackets.
Therefore, EF and JF are mathematically equivalent via
a Hamiltonian canonical transformation on the reduced phase space obtained with gauge fixing. 
The Weyl (conformal) Hamiltonian transformation maps solutions
of the equations of motion in the JF into solutions of the equations of motion in the EF
frame. 
We have shown examples of spherical symmetric static solutions in both frames. 
Since the Weyl (conformal) transformation connecting JF and EF is Hamiltonian canonical after gauge fixing,  the FJNW solution in the EF can be mapped into the BBMB black hole solution in the EF. 
The singularities in the transformation, connecting different frames, have been discussed in detail (see also 
\cite{Bronnikov:2002kf,Faraoni:2015paa,Kamenshchik:2016gcy,Domenech:2019syf,Kamenshchik:2024rkk}).
In fact, we looked, in particular, at the curvature invariants for the two static solutions in Section \ref{SubSect:sing}. In the EF the FJNW solution has two singular points ($r=0$ and $r=b$). Otherwise, the BBMB solution in the EF obtained using the Weyl (conformal) transformation has only one singular point ($\rho=0$). 
This difference in the properties of the two spacetimes is due to the fact that the transformation between the two frames is singular for a particular value of the radial coordinate ($r=b$ or $\rho=b/2$).
This is a further example of how the transformation from JF to EF can be seen as the generator of new solutions of the equations of motion. 
In fact, FJNW and BBMB are two non-equivalent physical solutions.
We believe that the physical equivalence between the frames connected by Weyl (conformal) transformation is still an open question. 
For instance, the effect of singularities on physical observables has to be carefully discussed in future works.

\begin{acknowledgement}
We thank Massimo Bianchi, Giampiero Esposito, Roberto Percacci  and an anonymous Referee for useful comments on this work.
\end{acknowledgement}

\appendix

\section{Equivalence between two decomposition of the trace of the Ricci tensor}
\label{App:1}
We discuss here the equivalence of two $3+1$ decompositions of the trace of the Ricci tensor.

In Eq. \eqref{covo} we used \cite{Hawking:1996ww}:
\begin{equation}
{}^{(4)}R={}^{(3)}R+K_{ij}K^{ij}-K^{2}+2\nabla_{\mu}\left(-n^{\mu}K-a^{\mu} \right)\, ,
\end{equation}
where $n^{\mu}$ is the unit time-like normal to the space-like surface $\Sigma$, and $a^{\mu}\equiv n^{\nu} \nabla_{\nu}n^{\mu}$.

In Eq. \eqref{ariscompongo} we used \cite{DeWitt1967}:
\begin{eqnarray}
\sqrt{-g}{}^{(4)}R&=&\sqrt{h}N\left({}^{3}R+K_{ij}K^{ij}-K^{2}\right)\nonumber\\
&&-(2K\sqrt{h}),_{0}+2f^{i},_{i}\,, 
\end{eqnarray}
where $f^{i}$ is defined as: 
\begin{equation}
f^{i}\equiv \sqrt{h}(KN^{i}-h^{ij}N,_{j})\,.
\end{equation}

The two decomposition are equivalent if: 
\begin{equation}
\nabla_{\mu} \left( -n^{\mu}K-a^{\mu}\right)=\frac{1}{\sqrt{-g}}\left(-(K\sqrt{h}),_{0}+f^{i},_{i} \right)\,.
\label{idento}
\end{equation}
First we notice that: 
\begin{eqnarray}
&&\nabla_{\mu}\left(-g^{\mu\nu}n_\nu K \right)=\nabla_{\mu}\left(-g^{\mu 0}n_0  K \right)\nonumber\\
&&=-\frac{1}{\sqrt{-g}}\partial_0\left(\sqrt{h}K\right)+\frac{1}{\sqrt{-g}}\partial_i\left(N^{i}\sqrt{h}K \right)\,.
\label{tento}
\end{eqnarray}
As regards the second term on the left side of \eqref{idento}, we can write:
\begin{eqnarray}
\nabla_{\mu}\left( a^{\mu}\right)&=&\nabla_{\mu}\left( g^{\mu\nu} n^{\rho}\nabla_{\rho}n_{\nu}\right)\nonumber\\
&=&\nabla_{\mu}\left( (h^{\mu\nu}-n^{\mu}n^{\nu}) n^{\rho}\nabla_{\rho}n_{\nu}\right)\nonumber\\
&=&\nabla_{i}\left(h^{ij}n^{\rho}\nabla_{\rho}n_{j} \right).
\label{tento1}
\end{eqnarray}
It is not difficult to see that: 
\begin{equation}
n^{\rho}\nabla_{\rho}n_{j}=-g^{\rho\, 0} n_0\Gamma^{0}_{\rho j}n_0=\frac{\partial_i N}{N}\,.
\label{calc}
\end{equation}
Therefore, at this point, we can write finally:
\begin{equation}
\nabla_{\mu}\left( a^{\mu}\right)=\frac{1}{\sqrt{-g}}\nabla_{i}\left(\sqrt{h} h^{ij}\partial_j N\right)\,.
\end{equation}

Using decomposition of the trace of the Ricci tensor defined here in Eq. $\eqref{ariscompongo}$, the action \eqref{aritutta} 
can be  re-written as: 
\begin{eqnarray}
S&=&\frac{1}{16\pi }
\int dt \int_{\Sigma_t}d^{3}x
\left(\sqrt{h}N({}^{(3)}R+K_{ij}K^{ij}-K^{2})\right)\nonumber\\
&&-\frac{1}{8\pi  }\int_{B}d^{3}x\sqrt{-\gamma}\,\left(\Theta+{}^{(3)}r_i n^{i} K -N\,{}^{(3)}r_i h^{ij}\nabla_{j}n^{0} \right) \nonumber\\
&&-\frac{1}{8\pi }\int_{B_{t_0}} d^2x \sqrt{\sigma} \, \mathrm{arcsinh} (n^\mu u_\mu) \nonumber\\
&& +\frac{1}{8\pi }\int_{B_{t_1}} d^2x \sqrt{\sigma} \,\mathrm{arcsinh} (n^\mu u_\mu), 
 \label{aritutta2}
\end{eqnarray}
and, also in with this decomposition, in the second line of the last formula the quantity in round parenthesis is proportional to the extrinsic curvature ${}^{(2)}k$ on two-dimensional boundary $B_t$ \cite{Hawking:1996ww}.  

%
\bibliographystyle{ieeetr}
\bibliography{bransdickepartcase}
%

\end{document}